\documentclass[5p]{elsarticle}

\usepackage{amsmath}
\usepackage{feynmf}
\usepackage{latexsym}
\usepackage{amssymb}
\usepackage{graphicx}
\usepackage{rotating}
\usepackage{latexsym}
\usepackage{verbatim}

 \def\be{\begin{equation}}
 \def\ee{\end{equation}}
 \def\bea{\begin{eqnarray}}
 \def\eea{\end{eqnarray}}
 \def\bei{\begin{itemize}}
 \def\eei{\end{itemize}}
 \def\bs{\begin{slide}}
 \def\es{\end{slide}}

 \def\L{\mathcal{L}}

 \def\g{\gamma}
 \def\G{\Gamma}
 
 \def\D{\Delta}
 \def\m{\mu}
 \def\n{\nu}

 \def\MZp{M_{Z'}}

 \def\({\left(}
 \def\){\right)}
 \def\[{\left[}
 \def\]{\right]}

 \def\cA{\mathcal{A}}

 \def\ds#1{#1\kern-1ex\hbox{/}}
\def\sla{\raise.15ex\hbox{$/$}\kern-.57em}

\begin{document}

\begin{frontmatter}

\title{A Phenomenological study on the wino radiative decay in anomalous $U(1)'$ models}

\author[label1]{Francesco Fucito}
\author[label1]{Andrea Lionetto}
\author[label2]{Antonio Racioppi}
\author[label1]{Daniel Ricci Pacifici}

\address[label1]{Dipartimento di Fisica dell'Universit\`a di Roma ,``Tor Vergata" and
I.N.F.N.~ -~ Sezione di Roma ~ ``Tor Vergata'',\\
 Via della Ricerca  Scientifica, 1 - 00133 ~ Roma,~ ITALY}
\address[label2]{National Institute of Chemical Physics and Biophysics, Ravala 10, Tallinn 10143, Estonia}

\begin{abstract}
An extension of the Standard Model by at least one extra U(1) gauge symmetry has been investigated by many authors.
In this paper we explore the possibility that this extra U(1) is anomalous. One of the possible signatures of this model
could be given by the photons produced in the decays of the NLSP into the LSP.
\end{abstract}


\end{frontmatter}

\section{Introduction}
The start of the LHC has greatly motivated detailed phenomenological studies of scenarios which involve physics beyond the Standard Model (SM).
Among them D-brane constructions in string theory
are one of the most promising framework in which the SM can be embedded and extended. Such brane constructions naturally lead to extra anomalous $U(1)$'s in
the four dimensional low energy theory and, in turn, to the presence of
possible heavy $Z^\prime$ particles
in the spectrum. These particles should be among the early findings of LHC
and besides for the above cited models they are also a prediction of
many other theoretical models of the
unification of forces (see \cite{Langacker:2008yv} for a recent review).
In~\cite{Anastasopoulos:2008jt} we have considered a minimal extension of the Minimal Supersymmetric Standard Model (MSSM) with a single extra $U(1)'$ gauge symmetry in a string-inspired setup.
We believe that our model encodes the key features of the low-energy sector of some of those
brane construction.
In this framework we studied in~\cite{Lionetto:2009dp} the radiative decay of the next to lightest supersymmetric particle (NLSP) into the lightest supersymmetric particle (LSP). This kind of process is very interesting since it might be the first one where the LSP could be observed at LHC~\cite{Baer:2002kv,Baer:2008ih} and at the upcoming ILC~\cite{Dreiner:2006sb,Basu:2007ys}.

\section{Preliminaries and Lagrangian}
Under suitable assumptions the LSP in our model turns out to be an axino~\cite{Fucito:2008ai}, the fermion component of the St\"uckelberg
supermultiplet related to the anomaly cancellation mechanism (see for details~\cite{Anastasopoulos:2008jt,Lionetto:2009dp,Fucito:2008ai}).
Without loss of generality we assume a wino-like NLSP.
In the following we just give the interaction term which involve the axino and the wino relevant for our analysis.
The interaction term, written in terms of four
components Majorana spinors\footnote{The gamma matrices $\gamma^\mu$ are
in the
Weyl representation.}, is given by
\bea
 && \!\!\!\!\!\!\!\!\!\!\!\!\!\!\!\!\!\!\!\!\!\!\!\!\!
 i \L =\sqrt{2} \sin\theta_W \frac{g_0 \cA^{(2)}}{\MZp}  \frac{g_2^2}{32 \pi^2 } \bar\lambda_2 \g_5 [\g^\m,\g^\n](\partial_\m A_\n) \psi_S
\eea
where $\lambda_2$ is the neutral wino, $\psi_S$ is the axino, $A_\n$ is the photon, $\theta_W$ the Weinberg angle,
$g_0$ and $g_2$ respectively the $U(1)'$ and $SU(2)$ coupling constants, $\cA^{(2)}$ the $U(1)'-SU(2)-SU(2)$ anomaly factor and $\MZp$ the $Z'$ mass.
The rate of the radiative decay ($\lambda_2 \to \Psi_S \g $) is
\be
 \G_\g^{(2)} =  g_2^4 \sin^2\theta_W \( \frac{g_0 \cA^{(2)}}{\MZp} \)^2   \frac{(\D M)^3 (\D M+2 M_S)^3}{1024 \pi^5 (\D M +M_S)^3} \!
\label{Gamma}
\ee
where $\D M=M_2-M_S$, while $M_2$ and $M_S$ are respectively the wino and axino masses.
As we showed in \cite{Lionetto:2009dp}, the radiative decay is the most dominant wino decay mode with a BR close to 1 ($\gtrsim 94\%$),
so we can use (\ref{Gamma}) to give an estimation of the wino mean life time
\be
 \tau_{\lambda_{2}} \simeq \frac{\hslash}{\G_\g^{(2)}}
\label{meanlife}
\ee

\section{LHC Phenomenology}
In order to fall into the WMAP range in the most experimentally attractive situation,
we considered a light LSP ($115 \, \text{GeV} \lesssim M_S \lesssim 150 \, \text{GeV}$) and a mass gap of order ${\D M}/{M_S} \simeq 20 \%$,
which imply more energetic  and therefore easier to detect photons. This requirement is necessary because the detector resolution increases with energy, while at low energy there is an obstruction for the detection of photons due to bremsstrahlung, QCD background  and absorption before the
detection from the calorimeter \cite{Aad:2009wy}.\\
Moreover we considered a universal squark mass $M_{\tilde Q}$ for the first two squark generations
(since under this assumption they are nearly degenerate) and we assumed flavor blindness~\cite{Baerbook}.
The contribution from the third generation squarks is always negligible.
\begin{figure}[t]
\centering
\includegraphics[scale=0.3]{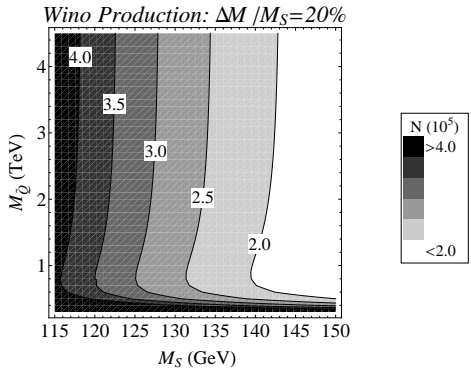}
      \caption{Number of directly produced winos in function of the axino mass $M_S$ and the universal squark mass $M_{\tilde Q}$.}
\label{fig:winoprod}
\end{figure}
In Fig.~\ref{fig:winoprod} we summarize the results obtained in~\cite{Lionetto:2009dp} by plotting the number of directly produced winos as a function of $M_S$ and $M_{\tilde{Q}}$ having assumed 14 TeV of center of mass energy and 100 fb$^{-1}$ of integrated luminosity.
Since the BR is almost close to one this is also the number of photons in the final state.
The number of photons produced is of the order of $10^5$.
In our analysis we follow~\cite{Aad:2009wy},\cite{Prieur:2005xv}-\cite{Terwort:2009zz},
where the NLSP decay in the GMSB framework is controlled by the parameter $C_{grav}$. If the NLSP lifetime is not too long ($C_{grav}\sim 1$)
photons originate close to the primary interaction vertex (``prompt photons'').
In the case of large $C_{grav}$ and therefore long lived neutralinos the resulting photons are non-pointing.
From now on we fix the axino mass $M_{S}\simeq 124$ GeV and the universal squark mass $M_{\tilde Q} \simeq 3.5$ TeV.
In our framework the role of $C_{grav}$ is played by the ratio ${g_0 \cA^{(2)}}/{\MZp}$.
In the following we discuss two different cases: short lived NLSP and long lived one.

\subsection{Short life time}

\begin{figure}[t!]
\centering
\includegraphics[scale=0.3]{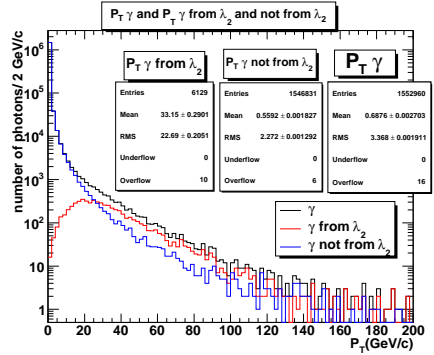}
\caption{$P_{T}$ distribution of photons (in log scale) for $10^4$ susy events.}
\label{fig:pta3}
\end{figure}

\begin{figure}[t!]
\centering
\includegraphics[scale=0.3]{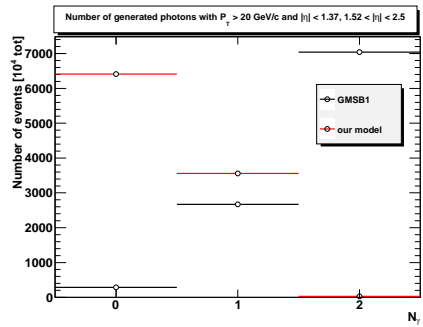}
\caption{Number of generated photons before the preselection cuts.}
\label{fig:num_phot_comp_before}
\end{figure}

\begin{figure}[t!]
\centering
\includegraphics[scale=0.3]{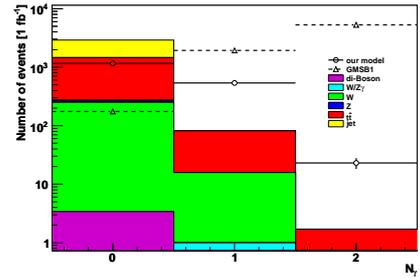}
\caption{Number of generated photons per event for our model and for GMSB1 and number of reconstructed photons for the SM
background, after having applied the cuts described in the main text.}
\label{fig:SMbackground}
\end{figure}

We compare the number of photons produced by radiative decay with the ones produced by the cascade decays of all the other supersymmetric processes.
We slightly modified the Herwig code 6.5~\cite{Corcella:2000bw} in order to take into account the new axino state in the neutral sector.
It should be stressed that Herwig does not implement extra $Z'$ in a supersymmetric framework.
This in turn implies that the total number of photons can be underestimated due to the lack of sparticles interactions with the $Z'$. However this problem can be overcome by assuming a decoupled $Z'$ either because it is very heavy or because it is extra-weak coupled.
We generated by Herwig 2-partons$\rightarrow$  2-sparticles events, using about 1 $fb^{-1}$ of integrated luminosity but we have not considered the case of SM particles produced directly in the parton-parton interaction.
A good discriminant variable of the process is the $P_{T}$ of the photons produced by radiative decay, in particular in the region of $P_{T}$ between 30-80 GeV/c.
The corresponding distribution is shown in Fig.~\ref{fig:pta3}.
We denote in red the number of $\gamma$'s radiatively produced from the decay of the wino, in blue the number of $\gamma$'s from all the other processes while in black the sum of the two.
We assumed $\tau_{\lambda_{2}} \simeq 1.29\cdot 10^{-15} s$, which is obtainable with $\MZp \simeq 1$ TeV and $g_0 \cA^{(2)} \simeq 0.2$.
We performed the same cut on the number of generated photons as in~\cite{Terwort:2009zz} with $P_{T}>20$ GeV and
with pseudorapidity $|\eta|\leq 1.37$, $1.52 < |\eta| < 2.5$,
which provides a good way to further suppress the SUSY background\footnote{After having employed the SUSY preselection cut which we describe later.}.
The result obtained by using Herwig in generating $10^{4}$ net events is given in Fig.~\ref{fig:num_phot_comp_before}.
The most important difference between our case and the GMSB1 sample~\cite{Aad:2009wy,Prieur:2005xv,Terwort:2008ii} is in the number of events with zero or two photons in the final state. The latter in particular is only 30 in our case.
This behavior can be related to the squark masses we have considered.
In our case they are about 3.5 TeV, while in the GMSB1 they are lower than 1 TeV ($\sim$ 900 GeV).
We choose the value of 3.5 TeV for the squark masses since in this case the number of directly produced winos essentially depends only on $M_S$ (see Fig.~\ref{fig:winoprod}).
The number of produced squarks is low since they have a high mass. Hence they give a lower contribution to the NLSP production.
If we consider lighter squarks, with masses less than 1 TeV, there is an increasing in the number of events with 2 photons in the final state.
In any case the channel with one photon in the final state is always the dominant one.\\
The key point in the analysis is the SM background discrimination.
We considered the same SM background as in~\cite{Terwort:2009zz}:
events with QCD jets, single gauge bosons (W and Z) production, di-Boson, $\gamma\gamma$ and $t\bar{t}$ production.
In order to disentangle the SM background from the signal we require a standard preselection cut for SUSY-like signatures:
\begin{itemize}
 \item[-] at least four jets must be present with $p_{T} > 50$ GeV ($p_{T} > 100$ GeV for the leading jet);
  \item[-] missing transverse energy $E^{miss}_{T}>60$ GeV.
\end{itemize}
After having applied these preselection cut we are able to reconstruct photons with $p_{T} > 20$ GeV and $|\eta|\leq 1.37$, $1.52 < |\eta| < 2.5$.
The jet-finder algorithm used in our analysis is the Durham-type clustering KTCLUS \cite{Catani:1993hr}.
The cut on the $E^{miss}_{T}$ requires some care.
In hadron colliders, the initial momentum of the colliding partons along the beam axis is not known
since the energy of each hadron is distributed and constantly exchanged between the partons. Hence the total amount of missing energy
cannot be determined in a straightforward way.
However the initial energy of particles travelling transverse to the beam axis is zero and thus any net momentum in the transverse
direction denotes missing transverse energy.
To determine the latter for the i-th not-detected particle in each generated event, we considered the following procedure: take
a transverse direction (perpendicular to the z axis) and define the vector
$$(\overrightarrow{E}^{miss}_T)_i=E_i\left(\frac{\overrightarrow{P}_T}{|\overrightarrow{P}|}\right)_i$$
whose direction is given by its momentum.
Then $E^{miss}_{T}=\sqrt{[(E^{miss}_{T})^{TOT}_x]^2+[(E^{miss}_{T})^{TOT}_y]^2}$ where
$(\overrightarrow{E}^{miss}_T)^{TOT}_{x}=\sum_i ((\overrightarrow{E}^{miss}_T)_{x})_i$ and analogously for the $y$ component.
The result is shown in Fig. \ref{fig:SMbackground}. In this plot the number of photons from the NLSP decay is that generated by Herwig while the number of background photons is the reconstructed one~\cite{Terwort:2009zz}.
We note that in the channels with $N_\gamma\geq 1$ the signal can be easily disentangled by the SM background.

\subsection{Long life time}
\begin{figure}[t!]
\centering
\includegraphics[scale=0.3]{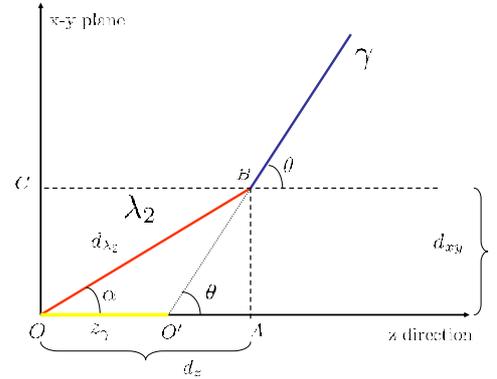}
\caption{Schematic diagram of a non-pointing photon. $d_{\lambda_{2}}$ is the distance traveled by $\lambda_{2}$ before its radiative decay. $z_{\gamma}$ is the value of the displaced vertex.}
\label{fig:nopoint}
\end{figure}
%
%
%
In the case in which the wino is a long lived particle the reconstruction of the emitted photon after its decaying plays an important role.
If the NLSP has a significant decay length, a photon will not ``point back" to the primary interaction point
($O$ in Fig. \ref{fig:nopoint}) but towards a kind of ``virtual'' point $(O')$.
The relevant discriminant variable is $z_{\gamma}$, the distance between $O$ and $O'$.
Using simple trigonometric relations we have:
\be
z_{\gamma} = d_{z}-d_{xy}\cot\theta
\ee
with $\cot\theta=P^{\gamma}_{Z}/P^{\gamma}_{T}$,
the ratio between the photon momentum along the beam direction (z-axis) and the transverse momentum.

\begin{figure}[t!]
\centering
\includegraphics[scale=0.3]{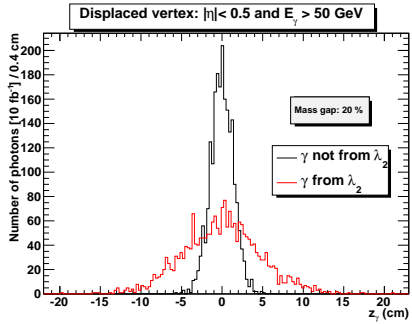}
\caption{Displaced vertex $z_{\gamma}$ for photons of $E>50$ GeV and  $\left|\eta\right|<0.5$: the red line stands for photons produced by wino decay, the black one for photons from the other susy processes.}
\label{fig:displaced}
\end{figure}

%
%
%
%
%
%
By fitting the vertex resolution along z as a function of the photon energy for $|\eta|<0.5$ (see plot in section ``Electrons and Photons" in~\cite{Aad:2009wy}) we derive the functional dependence of the standard deviation $\sigma$ for the Gaussian distribution associated to $z_{\gamma}$ on the energy in GeV of all the photons as
\be
 \frac{\sigma}{\text{mm}} \simeq 0.7 \sqrt{E/\text{GeV}}+\frac{65}{\sqrt{E/\text{GeV}}}
\label{eq:sigma}
\ee
To each $z_{\gamma}$ we can associate a random number sampled from a Gaussian distribution centered on that specific value of $z_{\gamma}$ and with $\sigma$
given by the relation in (\ref{eq:sigma}).
We considered a wino mean life time $\tau_{\lambda_{2}}\sim 3.463\cdot 10^{-10}$ s,
so that the distance travelled by this particle before its radiative decay is always within the ATLAS tracker radius ($\sim$ 115 cm).
In Fig. \ref{fig:displaced}  we show the results of the analysis for photons of $E>50$ GeV and with $\left|\eta\right| <0.5$
(photons with great angle to the beam direction, because these are more easily detected than the ones along the z direction),
for an integrated luminosity of 10 $fb^{-1}$.
By considering the value of their energy and the contribution of the tail of the associated Gaussian distribution, we observe a consequent enlargement of the distribution of $\gamma$'s not coming from $\lambda_2$, which otherwise would be a spike centered on the origin of the x-axis in Fig.~\ref{fig:displaced}.
The red line stands for photons produced by radiative wino decay, while the black one from the others SUSY processes.
We can see that in the region $|z_{\gamma}|>3$ cm the photons from the radiative decay are well distinguished from the background ones.
Since we have assumed $\Delta M/M_{S}=20\%$, $M_S\simeq 124$ GeV and $\tau_{\lambda_{2}}\sim 3.463\cdot 10^{-10}$ s,
we get from (\ref{meanlife})
\be
 \frac{g_0 \cA^{(2)}}{\MZp} \simeq \frac{3.9\cdot 10^{-4}}{\text{TeV}}
\label{ultima}
\ee
To satisfy (\ref{ultima}) we need either a super-heavy $Z'$ ($\MZp \simeq 385$ TeV if $g_0 \cA^{(2)}\simeq0.15$)
or an extra-weakly coupled ($g_0 \cA^{(2)}\simeq0.00039$ if $\MZp \simeq 1$ TeV) one.
This decoupling makes our simulation more consistent since in it we have neglected the $Z'$ contribution.
Cases of extra-weak $Z'$ were already studied in the past \cite{Langacker:2008yv,Feldman:2007wj}.

\section{Conclusion}
The most important result we get is shown in Fig.~\ref{fig:displaced}
which gives a distinctive behavior of the photons coming from our model with respect to those coming from other SUSY processes.
The distribution of $z_{\gamma}$ is centered in zero while the half-width is a function of the NLSP decay length $d_{\lambda_{2}}$. This means that the radiative produced photons are preferably emitted along the NLSP direction or more generally at very small angle. Thus the signal can be disentangled from the background by applying a suitable cut in a region $\left| z_{\gamma}\right| \lesssim 5$ cm.
Our result can be compared with~\cite{Aad:2009wy} in which the NLSP is a long lived wino-like particle in the particular gauge mediated model dubbed GMSB3. The main difference is in the mass gap.
In our model we considered a mass gap of $20\%$ and an LSP of 124 GeV, while in the GMSB3 case the mass gap is much higher: the gravitino (the LSP) is almost massless ($m_{\tilde{G}}=1.08\cdot 10^{-8}$ GeV) and the NLSP has a mass $m_{\tilde{\chi}_1}=118.8$ GeV.
In this case the energy distribution for the photons produced by the NLSP decay is peaked on the value of the NLSP mass $E\simeq m_{\tilde{\chi}_1}=118.8$ GeV while in our case the peak is around $20$ GeV. Moreover in the GMSB3 $c\tau_{\tilde{\chi_{1}}} \simeq 3.2$ m and so the NLSP long lived particles tend to escape the detector before their decay. This implies a sizable reduction in the number of photons detected.

\begin{flushleft}
{\bf Acknowledgments}
\end{flushleft}

\noindent The authors gratefully acknowledge the ATLAS group of Tor Vergata, in particular Prof.
A. Di Ciaccio, G. Cattani and R. Di Nardo for many stimulating discussions and help.
 A. R. would like to thank Prof. M. Raidal and Dr. K. Kannike for discussions and
the ESF JD164 contract for financial support.

\end{document}